\documentclass[pra,twocolumn,showpacs,preprintnumbers,amsmath,amssymb]{revtex4}

\usepackage{natbib}
\usepackage{graphicx}
\usepackage{dcolumn}
\usepackage{bm}
\newcommand{\bra}[1]{\langle#1|}

\newcommand{\ket}[1]{|#1\rangle}

\begin{document}

\preprint{APS/123-QED}

\title{Entanglement dynamics of two independent qubits\\ in environments with and without memory}

\author{B. Bellomo, R. Lo Franco, and G. Compagno}

\affiliation{%
Dipartimento di Scienze Fisiche ed Astronomiche, Universit\`{a} di
Palermo, via Archirafi 36, 90123 Palermo, Italy}

\date{\today}

\begin{abstract}
A procedure to obtain the dynamics of $N$ independent qudits ($d$-level systems) each interacting with its own reservoir, for any arbitrary initial state, is presented. This is then applied to study the dynamics of the entanglement of two qubits, initially in an extended Werner-like mixed state with each of them in a zero temperature non-Markovian environment. The dependence of the entanglement dynamics on the purity and degree of entanglement of the initial states and on the amount of non-Markovianity is also given. This extends the previous work about non-Markovian effects on the two-qubit entanglement dynamics for initial Bell-like states [B. Bellomo \textit{et al.}, Phys. Rev. Lett. \textbf{99}, 160502 (2007)]. The effect of temperature on the two-qubit entanglement dynamics in a Markovian environment is finally obtained.
\end{abstract}


\maketitle

\section{\label{intro}Introduction}
One of the most striking characteristics of quantum mechanics is represented by the entanglement. It plays a role in fundamental aspects, like non-locality \cite{bell,bennett}, and applicative ones, like quantum information processing \cite{niels}. It is therefore of interest the experimental generation and manipulation of entangled systems \cite{rai,polzik,blinov,haroche} and the theoretical study of the entanglement evolution. It is an important subject the analysis of entanglement decay and its relation with decoherence induced by the unavoidable interaction of the systems of interest with the environment.

The study of the entanglement evolution of two qubits, initially in a Bell state and exposed to local zero temperature and Markovian noisy environments, has led to the surprising result that it may markedly differ from the single qubit coherence evolution. In particular, in front of an exponential decrease of the single qubit coherence, the entanglement may instead completely disappear at a finite time \cite{diosi,yu1}, a phenomenon known as "entanglement sudden death" (ESD) which has been proven to occur in a quantum optics experiment \cite{almeida}. ESD appears to put a limitation to the time when entanglement could usefully exploited. Its intrinsic and practical interest has led to several investigations in Markovian environments at either zero or finite temperature and for initially pure and mixed states \cite{yu2,yu5,santos,tanas,sun,hamdou,carv1,cirone}. The disentanglement phenomenon being linked to the creation of body-environment correlations makes interesting to analyze its behavior in a non-Markovian system where memory effects are relevant \cite{ban1,ban2,kuanliu,yon,glendinning}.

Recently, a procedure has been developed that allows to obtain the complete dynamics of the two-qubit system by knowing the dynamics of each qubit interacting with its environment, which is applicable to any initial condition. This has permitted to show, for two noninteracting qubits initially described by pure Bell-like states and in non-Markovian environment at zero temperature, that entanglement may reappear after some period of time of complete disappearance \cite{bellomo}. This appears relevant from a fundamental point of view and because also may permit to extend the time of usefulness of entanglement.

The aim of this paper is to generalize the results obtained in Ref.~\cite{bellomo} first giving a formal extension of the procedure described above from two qubits to $N$ independent qudits ($d$-level systems), each interacting with its own reservoir. Then, we shall apply it to determine the entanglement dynamics of two identical qubits, initially described by a generalization of mixed Werner-like states, in non-Markovian environment at zero temperature, extending to a wider class of initial conditions the previous study. This will permit to find how entanglement dynamics and its revivals are related to physical parameters like the purity and the degree of entanglement of the initial states or to the amount of non-Markovianity of the system. Then, we shall exploit our procedure to extend in a simple way the previous analysis of two-qubit entanglement in a Markovian environment at finite temperature to the initial extended Werner-like states. This will permit to link the occurrence time of ESD to the purity or the degree of entanglement of the initial states.

The paper is organized as follows. In Section~\ref{procedure} we propose a general procedure to solve the dynamics of $N$ independent qudits. In Section~\ref{model} we give a standard Hamiltonian model describing the dynamics of single
qubit-reservoir interaction, and we also determine the two-qubit density matrix evolution. In Section~\ref{istatesandconcurrence} we introduce our initial states and recall the definition of concurrence to quantify the two-qubit entanglement. In Section~\ref{non-markoviandynamics} we study the two-qubit entanglement dynamics in non-Markovian environments at zero temperature while in Section~\ref{markoviandynamics} we analyze the effects of the temperature on the entanglement dynamics in Markovian environments. In Section~\ref{concl} we summarize our conclusions.

\section{\label{procedure}Procedure}
We consider a system formed by $N$ non-interacting parts
$\tilde{S}=\tilde{1},\tilde{2},\ldots,\tilde{N}$, each part
consisting of a $d$ level system (qudit) $S=1,2,\ldots,N$, locally
interacting respectively with a reservoir $R_S=R_1,R_2,\ldots,R_N$.
Each qudit $S$ and the corresponding reservoir $R_S$ are initially considered
independent. The evolution of the reduced density matrix of
the single qudit $S$ is given by
\begin{equation}\label{reducedrho}
\hat{\rho}^{S}(t)=\textrm{Tr}_{R_S}\left\{\hat{U}^{\tilde{S}}(t)\hat{\rho}^{S}(0)
\otimes\hat{\rho}^{R_S}(0)\hat{U}^{\tilde{S}\dag}(t)\right\},
\end{equation}
where the trace is over the reservoir $R_S$ degrees of freedom and
$\hat{U}^{\tilde{S}}(t)$ is the time evolution operator for the part
$\tilde{S}$. By using the spectral decomposition of the density
matrix $\hat{\rho}^{R_S}(0)$ of the reservoir
\begin{equation}\label{spectral decomposition}
   \hat{\rho}^{R_S}(0)= \sum_{\alpha_S}
   \lambda_{\alpha_S}\ket{\varphi_{\alpha_S}}\bra{\varphi_{\alpha_S}},
\end{equation}
the reduced density matrix $\hat{\rho}^{S}(t)$ can be expressed in the so called operator-sum
representation as \cite{niels}
\begin{equation}\label{krausrep}
\hat{\rho}^{S}(t)=\sum_{\alpha_S\beta_S}\hat{W}_{\alpha_S\beta_S}^{S}(t)\hat{\rho}^{S}(0)\hat{W}_{\alpha_S\beta_S}^{\dag
\,S}(t),
\end{equation}
where the operators $\hat{W}_{\alpha_S,\beta_S}^{S}(t)$ are given by
\begin{equation}\label{kraus operators}
    \hat{W}_{\alpha_S\beta_S}^{S}=\sqrt{\lambda_{\alpha_S}}\bra{\varphi_{\beta_S}}\hat{U}^{\tilde{S}}(t)\ket{\varphi_{\alpha_S}}.
\end{equation}

The assumption of independent parts implies that the time evolution
operator $\hat{U}(t)$ of the complete system
$\tilde{1}+\tilde{2}+\ldots+\tilde{N}$ factorizes as
$\hat{U}(t)=\hat{U}^{\tilde{1}}(t)\otimes
\hat{U}^{\tilde{2}}(t)\otimes \ldots \otimes \hat{U}^{\tilde{N}}(t)
$. It follows that the operator-sum representation of the reduced density
matrix for the $N$-qudit system ($1+2+\ldots+N$) reads like
{\setlength\arraycolsep{1.5pt}\begin{eqnarray}\label{kraustotal}
\hat{\rho}(t)&=&\sum_{\alpha_1\beta_1}\sum_{\alpha_2\beta_2}\cdots\sum_{\alpha_N\beta_N}
\hat{W}_{\alpha_1\beta_1}^{1}(t)\hat{W}_{\alpha_2\beta_2}^{2}(t)\cdots\hat{W}_{\alpha_N\beta_N}^{N}(t)\nonumber\\
&&\times\hat{\rho}^{T}(0)\hat{W}_{\alpha_1\beta_1}^{\dag
\,1}(t)\hat{W}_{\alpha_2\beta_2}^{\dag
\,2}(t)\cdots\hat{W}_{\alpha_N\beta_N}^{\dag \,N}(t).
\end{eqnarray}}
Given the basis $\{\ket{i_S},i_S=1_S,2_S,\ldots,d_S\}$ for the qudit $S$,
inserting the unity operators $I_S=\sum\ket{i_S}\bra{i_S}$ between operators and density matrix in
Eq.~({\ref{krausrep}}), it follows that the dynamics of each qudit
has the form
\begin{eqnarray}\label{singleevo}
\rho^{1}_{i_1i_1'}(t)&=&\sum_{l_1l_1'}A_{i_1i_1'}^{l_1l_1'}(t)\rho^{1}_{l_1l_1'}(0),
\nonumber \\
\,\rho^{2}_{i_2i_2'}(t)&=&\sum_{l_2l_2'}A_{i_2i_2'}^{l_2l_2'}(t)
\rho^{2}_{l_2l_2'}(0)\nonumber \\
& \vdots  & \nonumber \\
\rho^{N}_{i_Ni_N'}(t)&=&\sum_{l_Nl_N'}A_{i_Ni_N'}^{l_Nl_N'}(t)\rho^{N}_{l_Nl_N'}(0).
\end{eqnarray}
Adopting the same procedure for the $N$-qudit reduced density matrix $\hat{\rho}(t)$ given in
Eq.~({\ref{kraustotal}}), the dynamics of the $N$-qudit system is correspondingly expressed by
\begin{multline}\label{totalevo}
\rho_{i_1i_1',i_2i_2',\ldots,i_Ni_N'}(t)=\sum_{l_1l'_1}\sum_{l_2l'_2}\ldots\sum_{l_Nl'_N}A_{i_1i'_1}^{l_1l'_1}(t)
\\ A_{i_2i'_2}^{l_2l'_2}(t) \times \ldots \times
A_{i_N i'_N}^{l_Nl'_N}(t)
\rho_{l_1l'_1,l_2l'_2,\ldots,l_Nl'_N}(0),
\end{multline}
where $i_S=1_S,2_S,\ldots,d_S$ with $S=1,\ldots,N$.
Eqs.~(\ref{singleevo}) and (\ref{totalevo}) clearly show that the
dynamics of $N$-qudit density matrix elements follows by
knowing the ones of each single qudit, and that it is applicable to any initial condition of the total system.

We wish to point out that the procedure developed above, which permits to obtain the dynamics of $N$ independent qudits each locally interacting with its own reservoir, can be also applied to the case where the reservoir is unique but the qudits are placed at distances larger than the spatial correlation length of the reservoir. Another point we wish to stress is the practical aspect of our procedure. In fact, it allows to obtain the dynamics of $N$ qudits, provided that the dynamics of each qubit is known, by purely algebraic way and independently from the initial conditions. This considerably simplify the work typically required by writing and solving differential equations for the density matrix elements of the total system \cite{qasimi} or by using other approaches whose difficulty increases with the number of parts, levels or
environmental noise complexity of the total system \cite{carv1}.

\section{Model\label{model}}
We now shall apply the above procedure to the case of a
system composed by two parts, each one consisting of a two-level
system (qubit) interacting with a reservoir. We shall take
the single part ``qubit+reservoir'' described by the Hamiltonian
\begin{equation}\label{Hamiltonian}
\hat{H}=\hbar \omega_0 \hat{\sigma}_+\hat{\sigma}_-+\sum_k \left[
\hbar \omega_k \hat{b}_k^\dag \hat{b}_k+\left(g_k \hat{\sigma}_+
\hat{b}_k  + g_k^* \hat{\sigma}_- \hat{b}_k^\dag\right)\right],
\end{equation}
where $\omega_0$ is the transition frequency and $\sigma_ \pm$ are the system raising and lowering
operators of the qubit, while $b_k^\dag $, $b_k $ are the creation and annihilation operators and $g_k$ the coupling
constant of the mode $k$ with frequency $\omega_k$. The Hamiltonian of Eq.~(\ref{Hamiltonian}) is used to describe
various systems such as a qubit formed by an exciton in a potential well environment. However to fix our ideas we shall take
it to represent a qubit formed by the excited and ground electronic
state of a two-level atom interacting with the reservoir formed by
the quantized modes of a high-$Q$ cavity. The dynamics of an atom interacting with a quantum electromagnetic field
under the dipole, rotating-wave, and two-level approximation is in
fact described by the Hamiltonian of Eq.~(\ref{Hamiltonian}).

Systems described by this Hamiltonian have been studied under
different conditions, as weak coupling and finite temperature in both Markovian
\cite{barnbook} and non-Markovian \cite{shresta} approximation, and strong coupling at zero temperature
\cite{petru,maniscalco}. In all of these cases, as we shall report in
Sections~\ref{non-markoviandynamics} and \ref{markoviandynamics}, the single-qubit density matrix evolution can be cast in the following form:
\begin{equation}\label{roA}
\hat{\rho}^S(t)=\left(%
\begin{array}{cc}
\rho^S_{11}(t) &  \rho^S_{10}(t)\\\\
 \rho^S_{01}(t)  &  \rho^S_{00}(t)  \\
\end{array}\right),
\end{equation}
where
\begin{eqnarray}\label{singlequbitevo}
  \rho^S_{11}(t) &=& u^S_t \rho^S_{11}(0)+v^S_t \rho^S_{00}(0),\nonumber \\
  \rho^S_{00}(t) &=& (1-u^S_t) \rho^S_{11}(0)+(1-v^S_t) \rho^S_{00}(0), \nonumber\\
  \rho^S_{10}(t) &=&\rho^{S\ast}_{01}(t)=z^S_t  \rho^S_{10}(0),
\end{eqnarray}
$u^S_t,v^S_t,z^S_t$ being functions of the time $t$. Now we are
ready to use, following the procedure described before, the
evolution of the reduced density matrix elements for the single
qubit to construct the reduced density matrix $\hat{\rho}$ for the
two-qubit system. The two qubits, labeled by  $S=A,B$, may be in general in
different environments so that their evolution is characterized by different values of the
functions $u^S_t,v^S_t,z^S_t$. In the
standard basis $\mathcal{B}=\{\ket{1}\equiv\ket{11},
\ket{2}\equiv\ket{10}, \ket{3}\equiv\ket{01}, \ket{4}\equiv\ket{00}
\}$, using Eqs.~(\ref{singleevo}), (\ref{totalevo}) and (\ref{singlequbitevo}),
we obtain for the diagonal elements
\begin{eqnarray}\label{rototdiag}
\rho_{11}(t)&=&u^A_tu^B_t\rho_{11}(0)+u^A_t v^B_t \rho_{22}(0)\nonumber\\&+&v^A_tu^B_t\rho_{33}(0)+v^A_tv^B_t\rho_{44}(0),\nonumber\\
\rho_{22}(t)&=&u^A_t(1-u^B_t)\rho_{11}(0)+u^A_t(1-v^B_t)\rho_{22}(0)\nonumber\\&+ &v^A_t(1-u^B_t)\rho_{33}(0)+v^A_t(1-v^B_t)\rho_{44}(0),\nonumber\\
\rho_{33}(t)&=&(1-u^A_t)u^B_t \rho_{11}(0)+(1-u^A_t)v^B_t\rho_{22}(0)\nonumber \\&+&(1-v^A_t)u^B_t\rho_{33}(0)+(1-v^A_t)v^B_t\rho_{44}(0),\nonumber\\
\rho_{44}(t)&=&(1-u^A_t)(1-u^B_t)\rho_{11}(0)+(1-u^A_t)\nonumber\\
&&\times(1-v^B_t)\rho_{22}(0)+(1-v^A_t)(1-u^B_t)\rho_{33}(0)\nonumber\\&+&(1-v^A_t)(1-v^B_t)\rho_{44}(0),
\end{eqnarray}
and for the non-diagonal elements
\begin{eqnarray}\label{rototnodiag}
\rho_{12}(t)&=&u^A_tz^B_t\rho_{12}(0)+v^A_tz^B_t\rho_{34}(0) ,\nonumber\\
 \rho_{13}(t)&=&z^A_tu^B_t\rho_{13}(0)+z^A_tv^B_t\rho_{24}(0),\nonumber\\
\rho_{14}(t)&=& z^A_tz^B_t \rho_{14}(0),\quad \rho_{23}(t)= z^A_tz^{B\ast}_t\rho_{23}(0),\nonumber\\
\rho_{24}(t)&=&z^A_t (1-u^B_t)\rho_{13}(0) + z^A_t (1-v^B_t)\rho_{24}(0),\nonumber\\
\rho_{34}(t)&=& (1-u^A_t)z^B_t \rho_{12}(0) + (1-v^A_t) z^B_t
\rho_{34}(0),
\end{eqnarray}
with $\rho_{ij}(t)=\rho^*_{ji}(t)$, $\hat{\rho}(t)$ being a hermitian matrix. We point out that Eqs.~(\ref{rototdiag}) and (\ref{rototnodiag}), being valid for any form of the single-qubit evolution functions $u^S_t,v^S_t,z^S_t$, are more general of the corresponding ones given in Ref.~\cite{bellomo}, where the reduced density matrix elements have been given for the particular case of non-Markovian environment at zero temperature, and they also allow to study the two-qubit density matrix evolution for any initial state.

\section{\label{istatesandconcurrence}Initial states and concurrence}
The two-qubit entanglement dynamics has been previously analyzed taking as initial states pure Bell-like states \cite{bellomo}. However, as said before, it appears of interest to obtain the entanglement dynamics of two independent qubits, each locally interacting with a reservoir, in the case of more general initial conditions and in particular for Werner states \cite{yu5}. The procedure we have developed is applicable to any two-qubit initial state and therefore also to Werner states. In particular, in this section we consider as two-qubit initial states an extension of the Werner states. These ``extended'' Werner-like (EWL) states are mixed states that may reduce to Werner mixed states \cite{werner,munro,wei} or to Bell-like pure states. They shall allow us to put in evidence the effects of both the mixedness and the degree of entanglement of the initial states on the entanglement dynamics.

\subsection{\label{initial state}Initial states}
It is known that Bell states and Werner mixed states are important in quantum information and computation processing \cite{niels,popescu}. Werner-like mixed states are also shown to occur during an experimental generation of Bell states \cite{hag}. However, from Eqs.~(\ref{rototdiag}) and (\ref{rototnodiag}) it is evident that the two-qubit density matrix at time $t$ is valid for arbitrary initial states and its structure crucially depends on their form. We use this property to study the two-qubit entanglement dynamics starting from EWL states defined as
\begin{eqnarray}\label{a}
    \hat{\rho}^\Phi(0)&=&r \ket{\Phi}\bra{\Phi}+\frac{1-r}{4}I_4,\nonumber\\
    \hat{\rho}^\Psi(0)&=&r \ket{\Psi}\bra{\Psi}+\frac{1-r}{4}I_4.
\end{eqnarray}
In Eq.~(\ref{a}), $r$ indicates the purity of the initial states, $I_4$ is the $4\times4$ identity matrix and
\begin{eqnarray}
\ket{\Phi}=\alpha\ket{01}+\beta\ket{10},\quad\ket{\Psi}=\alpha\ket{00}+\beta\ket{11},\label{istates}
\end{eqnarray}
are the Bell-like states with $\alpha$ real, $\beta=|\beta|e^{i\delta}$ and $\alpha^2+|\beta|^2=1$. The states defined by Eq.~(\ref{a}) reduce to the well-known Werner-like states \cite{munro,wei} when their pure part becomes a Bell state, that is for $\alpha=\pm\beta=1/\sqrt{2}$. The degree of entanglement of the Werner-like states cannot be increased by any unitary transformation, so that they correspond to a maximum degree of entanglement for a given linear
entropy. For $r=0$ the Werner-like states become totally mixed states, while for $r=1$ they are the well-known Bell states. On the other hand, fixing $r=1$, the EWL states of Eq.~(\ref{a}) reduce respectively to the Bell-like pure states $\ket{\Phi},\ket{\Psi}$ of Eq.~(\ref{istates}). The EWL states $\hat{\rho}^\Phi(0),\hat{\rho}^\Psi(0)$ are therefore more general than the Werner-like states since they contain the possibility that their pure part is not maximally entangled. They, as said before, thus permit to study the effect of both the initial states mixedness and the degree of entanglement of their pure part on the entanglement dynamics. The initial density matrix elements are for the state $\hat{\rho}^\Phi$
\begin{eqnarray}\label{wernerstate1}
\rho^\Phi_{11}(0)&=&\frac{1-r}{4}, \quad
\rho^\Phi_{22}(0)=\frac{1-r}{4}+|\beta|^2 r,\nonumber\\
\rho^\Phi_{33}(0)&=&\frac{1-r}{4}+\alpha^2 r, \quad
\rho^\Phi_{44}(0)=\frac{1-r}{4} , \nonumber \\
\rho^\Phi_{14}(0)&=& 0,\quad \rho^\Phi_{23}(0)= \alpha \beta r,
\end{eqnarray}
and for the state $\hat{\rho}^\Psi$
\begin{eqnarray}\label{wernerstate2}
\rho^\Psi_{11}(0)&=&\frac{1-r}{4} +|\beta|^2 r, \quad
\rho^\Psi_{22}(0)=\frac{1-r}{4},\nonumber\\
\rho^\Psi_{33}(0)&=&\frac{1-r}{4}, \quad
\rho^\Psi_{44}(0)=\frac{1-r}{4}+\alpha^2 r , \nonumber \\
\rho^\Psi_{14}(0)&=& \alpha \beta r,\quad \rho^\Psi_{23}(0)= 0.
\end{eqnarray}
The structure of the density matrix elements of Eqs.~(\ref{wernerstate1}) and (\ref{wernerstate2}) show that the EWLSs belong to the class of the ``X'' states, that is to those states having non-zero elements only along the main diagonal and anti-diagonal. The general structure of an ``X'' density matrix is thus
\begin{equation}\label{Xstates}
   \hat{\rho} (0)= \left(
\begin{array}{cccc}
  \rho_{11}(0) & 0 & 0 & \rho_{14}(0)  \\
  0 & \rho_{22}(0) & \rho_{23}(0) & 0 \\
  0 & \rho_{23}(0)^* & \rho_{33}(0) & 0 \\
  \rho_{14}(0)^* & 0 & 0 & \rho_{44}(0) \\
\end{array}
\right).
\end{equation}
Bell states and Werner states belong to this class of states and we also point out that a ``X'' structure density matrix may also arise in a wide variety of physical situations \cite{bose2001,pratt,wang,hag}. A remarkable aspect of the ``X'' states is that, under the single qubit evolutions of Eq.~(\ref{singlequbitevo}) determined by the Hamiltonian of Eq.~(\ref{Hamiltonian}), the initial ``X'' structure is maintained during the evolution, as easily seen from Eqs.~(\ref{rototdiag}) and (\ref{rototnodiag}). Moreover, it is know that, for such a ``X'' density matrix, the expression of the concurrence in terms of the density matrix elements takes a particular simple form \cite{yu5}.

\subsection{\label{concurrence}Concurrence}
In order to describe the entanglement dynamics of the bipartite
system, we use Wootters concurrence \cite{wootters}. This is
obtained from the density matrix $\hat{\rho}$ of the two-qubit system $A+B$ as
\begin{equation}\label{concdefinition}
C_{\hat{\rho}}(t)=\mathrm{max}\{0,
\sqrt{\lambda_1}-\sqrt{\lambda_2}-\sqrt{\lambda_3}-\sqrt{\lambda_4}\},
\end{equation}
where the quantities $\lambda_i$ are the eigenvalues of the matrix
$\zeta =\hat{\rho} (\sigma_y^A \otimes \sigma_y^B)\hat{\rho}^*
(\sigma_y^A \otimes \sigma_y^B)$, arranged in decreasing order. Here $\sigma_y^S$ are the well-known Pauli matrices and
$\hat{\rho}^*$ denotes the complex conjugation of $\hat{\rho}$ in
the standard basis $\mathcal{B}$. $C_{\hat{\rho}}(t)$ varies from 0 for a
disentangled state to 1 for a maximally entangled state. In our case, $C_{\hat{\rho}}(t)$ as a function of the two-qubit density matrix elements of Eqs.~(\ref{rototdiag}) and (\ref{rototnodiag}) depends on the initial condition and on the dynamics as expressed by the functions $u_t^S,v_t^S,z_t^S$.

The concurrence at the time $t$ for an initial ``X'' state defined
in Eq.~(\ref{Xstates}) can be easily computed by exploiting the fact that, under our dynamical conditions, the ``X'' structure is maintained during the two-qubit evolution. It results to be given by \cite{yu5}
\begin{equation}\label{concurrence x state}
C_\rho^X(t)=2\mathrm{max}\{0,K_1(t),K_2(t)\},
\end{equation}
where
\begin{eqnarray}\label{concxstate}
K_1(t)=|\rho_{23}(t)|-\sqrt{\rho_{11}(t)\rho_{44}(t)}\nonumber \\
K_2(t)=|\rho_{14}(t)|-\sqrt{\rho_{22}(t)\rho_{33}(t)} &&.
\end{eqnarray}

The EWL states of Eq.~(\ref{a}) have the same initial value of the concurrences, given
by
\begin{equation}
C_\rho^{\Phi}(0)=C_\rho^{\Psi}(0)=2\mathrm{max}\{0,(\alpha|\beta|+1/4)r-1/4\},
\end{equation}
from which one finds that there is initially entanglement when $r>r^\ast=(1+4\alpha|\beta|)^{-1}$, thus entanglement exists for these states only if $r>1/3$. In particular, the Bell-like pure states $\ket{\Phi},\ket{\Psi}$, obtained by $\rho^{\Phi}(0),\rho^{\Psi}(0)$ for $r=1$, have the same degree of entanglement (concurrence) regulated by $\alpha$ according to $C_\Phi(0)=C_\Psi(0)=2\alpha\sqrt{1-\alpha^2}$.
Using the expression of the concurrence for ``X'' states given in Eqs.~(\ref{concurrence x state}) and (\ref{concxstate}), it is readily seen that $C_\rho^X(t)$ for the EWL states of Eq.~(\ref{a}) is respectively
\begin{equation}\label{concwstate}
C_\rho^\Phi(t)=2\mathrm{max}\{0,K_1(t)\},\quad
C_\rho^\Psi(t)=2\mathrm{max}\{0,K_2(t)\}.
\end{equation}
We shall use these equations in order to study the two-qubit entanglement dynamics. In particular, from now on we shall limit our analysis to the case of two identical qubits locally interacting with identical environments. This implies that the functions $u^S_t,v^S_t,z^S_t$ ($S=A,B$),
introduced in Eq.~(\ref{singlequbitevo}), are the same for both qubits and we shall then indicate them simply by $u_t,v_t,z_t$.

\section{\label{non-markoviandynamics}Non-Markovian entanglement dynamics}
In this section, we shall analyze the two-qubit entanglement dynamics for our EWLSs of Eq.~(\ref{a}) in the non-Markovian case at zero temperature. This constitutes an extension of our previous study of the non-Markovian effects on the two-qubit entanglement dynamics for the initial Bell-like pure states of Eq.~(\ref{istates}) \cite{bellomo}. This extension shall permit to point out how the two-qubit entanglement dynamics is affected by the presence of mixedness in the initial states regulated by the purity parameter $r$ and the degree of entanglement represented by $\alpha$. On the basis of our procedure described in Sec.~\ref{procedure}, we shall obtain the reduced two-qubit dynamics by the knowledge of the reduced single qubit dynamics.

Single qubit dynamics induced by the Hamiltonian of Eq.~(\ref{Hamiltonian}) in the non-Markovian limit has been treated in several works \cite{anasto,shresta,maniscalco}. In the following, we exploit the results there obtained about this
dynamics under weak coupling at finite (low) temperature and strong coupling at zero temperature.

\subsection{Weak coupling (low $T$)}
The exact single qubit dynamics induced by Hamiltonian of Eq.~(\ref{Hamiltonian}), with the field
initially at finite temperature, has been recently studied
\cite{shresta}, and a formal but non transparent, because of its complexity, solution for the density matrix
elements evolution has been obtained. However, at low temperature
($e^{-\hbar\omega_0/k_BT}\ll 1$) and weak coupling
($\Gamma/\omega_0\ll 1$, $\Gamma$ being the zero temperature spontaneous emission rate), a simple explicit solution has been found, that can be put in the form of Eq.~(\ref{singlequbitevo}) with
\begin{eqnarray}
u_t&=&1-\frac{(1-e^{-\Gamma t})(1-e^{-\frac{\hbar\omega_0}{k_BT}})}{(1-e^{-\frac{\hbar\omega_0}{k_BT}-\Gamma t})^2},\nonumber\\
v_t&=&\frac{e^{-\frac{\hbar\omega_0}{k_BT}}(1-e^{-\Gamma t})}{1-e^{-\frac{\hbar\omega_0}{k_BT}-\Gamma t}},\nonumber\\
z_t&=&\frac{1-e^{-\frac{\hbar\omega_0}{k_BT}}}{1-e^{-\frac{\hbar\omega_0}{k_BT}-\Gamma
t}}e^{-\Gamma t/2-i\omega_0t}.
\end{eqnarray}
This solution presents some differences with respect to the corresponding low temperature
Markovian dynamics, for example the upper level relaxation time results longer than the Markovian one \cite{shresta}. However, it is possible to show that the evolution of the concurrence for the two initial states $\hat{\rho}^{\Phi}(0),\hat{\rho}^{\Psi}(0)$ of Eq.~(\ref{a}) is not qualitatively much different from the Markovian one that we shall treat in Sec.\ref{markoviandynamics}. In the low temperature regime, a strong qubit-reservoir coupling condition seems to be required in order to have evident non-Markovian effects in
the entanglement dynamics of a two-qubit system. To this aim, in the following we shall then use the results for the non-Markovian strong coupling regime obtained at zero temperature \cite{maniscalco}.

\subsection{Strong coupling ($T=0$)}
Following our previous work \cite{bellomo}, here we use the exact solutions for the single qubit dynamics
at zero temperature \cite{petru,maniscalco} with the aim to extend the analysis of two-qubit entanglement dynamics to the more general initial conditions assigned by the EWLSs of Eq.~(\ref{a}).

Because the Hamiltonian of Eq.~(\ref{Hamiltonian}) represents a model for the damping of an atom in a cavity, in the case of a single excitation in the atom-cavity system, the effective spectral
density $J(\omega)$ is taken as the spectral distribution of an
electromagnetic field inside an imperfect cavity supporting the mode
$\omega_0$, resulting from the combination of the reservoir spectrum
and the system-reservoir coupling, with $\Gamma$ related to the
microscopic system-reservoir coupling constant. This spectral density has the form \cite{petru}
\begin{equation}\label{spectraldensity}
J(\omega)=\frac{1}{2 \pi}\frac{\Gamma
\lambda^2}{(\omega_0-\omega)^2+\lambda^2},
\end{equation}
where $\lambda$, defining the spectral width of the coupling, is
connected to the reservoir correlation time $\tau_B$ by the relation
$\tau_B \approx \lambda^{-1}$. In fact, it can be shown that the
reservoir correlation function, corresponding to the spectral
density of Eq.~(\ref{spectraldensity}), has an exponential form with
$\lambda$ as decay rate. On the other hand the parameter $\Gamma$
can be shown to be related to the decay of the excited state of the
atom in the Markovian limit of flat spectrum. The relaxation time
scale $\tau_R$ over which the state of the system changes is then
related to $\Gamma$ by $\tau_R \approx \Gamma^{-1}$. The exact
solution for the single-qubit density matrix at zero temperature
with the spectral density given by Eq.~(\ref{spectraldensity}) has
the form of Eq.~(\ref{singlequbitevo}) with \cite{petru,maniscalco}
\begin{eqnarray}\label{uvznonmark}
u_t&=&\mathrm{e}^{-\lambda t}\left[ \cos \left(\frac{dt}{2}\right)+\frac{\lambda}{d}\sin \left(\frac{dt}{2}\right)\right]^2,\nonumber\\
v_t&=&0,\nonumber\\
z_t&=&\sqrt{u_t},
\end{eqnarray}
where $d=\sqrt{2\Gamma \lambda-\lambda^2}$. A weak and a strong
coupling regime can be distinguished for the single qubit dynamics.
The weak regime corresponds to the case $\lambda/\Gamma>2$, that is
$\tau_R>2\tau_B$. In this regime the relaxation time is greater than
the reservoir correlation time and the dynamics is essentially
Markovian with an exponential decay controlled by $\Gamma$. In the
strong coupling regime, that is for $\lambda/\Gamma<2$, or $\tau_R<
2\tau_B$, the reservoir correlation time is greater or of the same
order of the relaxation time and non-Markovian effects become
relevant. Therefore, we shall study the two-qubit
entanglement dynamics in this regime. Substituting the expressions
for $u_t$, $v_t$ and $z_t$ given by Eq.~(\ref{uvznonmark}) in
Eqs.~(\ref{rototdiag}) and (\ref{rototnodiag}), the
two-qubit density matrix at time $t$ for both the initial states of
Eq.~(\ref{a}) is obtained. Using for the concurrence Eq.~(\ref{concwstate}),
we then determine its dependence on time $t$, purity
parameter $r$, probability amplitude $\alpha$ and
a dimensionless parameter $\lambda/\Gamma$ that indicates
the coupling regime. The explicit expressions of concurrence in terms of these parameters are, although simple, rather long and not particularly enlightening. Therefore, we shall plot the concurrence as a
function of both $ \Gamma t$ and one of the dimensionless parameters
($r,\alpha^2,\lambda/\Gamma$), after fixing the remaining ones.

\begin{figure}
\begin{center}
\includegraphics[width=6.25 cm, height=3.7 cm]{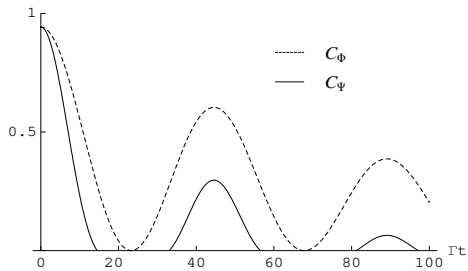}
\caption{\label{PRLcomparison}\footnotesize{(Non-Markovian case; $r=1$,
$\alpha^2=1/3$, $T=0$, $\lambda/\Gamma=0.01$) Comparison between concurrences
$C_\Phi(t)$ (dashed line) and $C_\Psi(t)$ (solid line) as a function of the dimensionless quantity $\Gamma t$.}}
\end{center}
\end{figure}
The case for $r=1$, that is when the initial EWL states of Eq.~(\ref{a}) reduce to the pure Bell-like states of Eq.~(\ref{istates}), has been previously treated \cite{bellomo}. The concurrence time evolution in terms of $\Gamma t$ and $\alpha^2$ has been analyzed, for the fixed value $\lambda/\Gamma=0.1$. This choice of $\lambda/\Gamma$ corresponds to an experimentally feasible value in the cavity QED context. In fact, cavity QED experimental configurations have been recently realized using Rydberg atoms with lifetimes $T_\textrm{at}\approx30\textrm{ms}$, inside Fabry-Perot cavities
with quality factors $Q\approx4.2\times10^{10}$ giving a cavity lifetime $T_\textrm{cav}\approx130\textrm{ms}$ \cite{kuhr},
these values corresponding to $\lambda/\Gamma\approx 0.1$. In particular, it has been shown that $C_\Phi(t)$, related to the initial Bell-like state $\hat{\rho}^{\Phi}(0)=\ket{\Phi}\bra{\Phi}$, periodically vanishes according to the
zeros of the function $u_t$ of Eq.~(\ref{uvznonmark}) with a damping of its revival
amplitude and ESD does not occur. The concurrence $C_\Psi(t)$, related to the initial state $\hat{\rho}^{\Psi}(0)=\ket{\Psi}\bra{\Psi}$, has been shown to have a similar behavior for values of $\alpha^2\geq1/2$, while for $\alpha^2<1/2$ two ranges of the parameter may be distinguished, one where ESD occurs and another where revival of entanglement appears after periods of times (dark periods) when entanglement has completely disappeared. As an example of this behavior, we plot in Fig.~\ref{PRLcomparison} the concurrences $C_\Phi(t)$ and $C_\Psi(t)$ for $\alpha^2=1/3$ and $\lambda/\Gamma=0.01$ which corresponds to a strong non-Markovian coupling. This choice of $\lambda/\Gamma$ is taken to make the revival phenomenon evident. Fig.~\ref{PRLcomparison} shows that the remarkable phenomenon of entanglement revivals occur, for the state $\hat{\rho}^{\Psi}(0)=\ket{\Psi}\bra{\Psi}$, after finite dark periods. Moreover, we wish to point out that, for initial Bell states, the concurrences at the time $t$ take the simple form
\begin{equation}\label{concurrenceBell}
C_\Phi(t)=u_t,\quad
C_\Psi(t)=u^2_t,\quad(r=1,\alpha=|\beta|=1/\sqrt{2})
\end{equation}
with the zeros of the function $u_t$ of Eq.~(\ref{uvznonmark}) given by
\begin{equation}
t_n=\frac{2}{d}[n\pi-\arctan(d/\lambda)].
\end{equation}
The distance $\Delta t$ between two consecutive zeros of $u_t$ and then of the concurrences of Eq.~(\ref{concurrenceBell}), is constant, depending only on the ratio $\lambda/\Gamma$ as
\begin{equation}
\Delta t=\frac{2\pi/\Gamma}{\sqrt{\lambda/\Gamma(2-\lambda/\Gamma)}}.
\end{equation}

\begin{figure}
\begin{center}
\includegraphics[width=4.25 cm, height=3.7 cm]{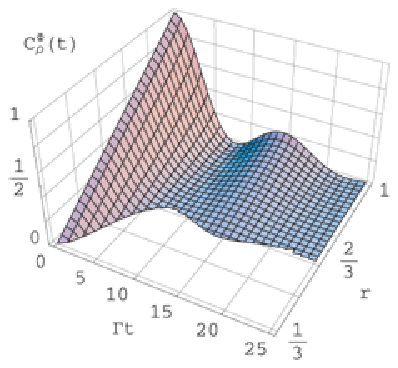}
\includegraphics[width=4.25 cm, height=3.7 cm]{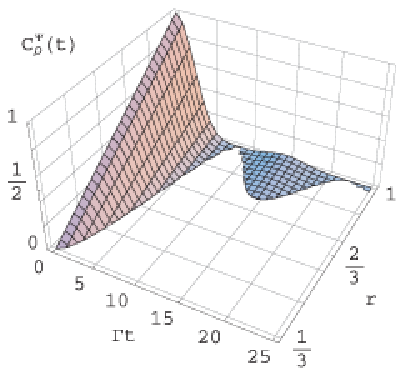}
\caption{\label{nc(r,t)}\footnotesize{(Non-Markovian case; $T=0$,
$\alpha^2=|\beta|^2=1/2$, $\lambda/ \Gamma=0.1$) Concurrence
$C_\rho^{\Phi}(t)$ (left figure) and $C_\rho^{\Psi}(t)$ (right
figure) for Werner-like states as a function of the dimensionless quantities $\Gamma t$ and $r$.}}
\end{center}
\end{figure}
Because our procedure can be easily applied to any initial state, we extend the analysis of the two-qubit entanglement dynamics for non-Markovian environments from pure Bell-like states to the interesting case of initial mixed states \cite{yu5} of the EWL form given by Eq.~(\ref{a}). Therefore, the new aspects we wish to examine here correspond to the case when the purity $r$ of the initial EWL states of Eq.~(\ref{a}) is less than one. In particular, the concurrences $C_\rho^{\Phi}(t)$ and $C_\rho^{\Psi}(t)$ are plotted in Fig.~\ref{nc(r,t)} in terms of the dimensionless quantities $\Gamma t$ and $r$, for $\lambda/\Gamma=0.1$ and $\alpha=1/\sqrt{2}$. This choice of $\alpha$ corresponds to take Werner-like states as initial states, as described in Sec.~\ref{initial state}. Different choices of $\alpha$ values do not give entanglement dynamics qualitatively different from the case treated here. From Fig.~\ref{nc(r,t)} we see that ESD occurs for both the initial Werner-like states $\hat{\rho}^\Phi(0)$ and $\hat{\rho}^\Psi(0)$ and sufficiently small values of $r$. This is at variance with the behavior of initial Bell-like states, where ESD occurred only for the state $\hat{\rho}^{\Psi}(0)=\ket{\Psi}\bra{\Psi}$. For intermediate values of purity $r$, entanglement revivals after finite dark periods occur for both initial Werner-like states. This again differs from previous results \cite{bellomo} where revivals and dark periods of entanglement occurred only for the Bell-like state $\ket{\Psi}$. In the region of $r$ large ($r\rightarrow1$) the time behavior of the concurrences approaches the one of initial Bell states. The reason of the decrease of the concurrences value at $t=0$ with the decrease of the purity $r$ is linked to the corresponding increase of the initial states mixedness. On the other hand, the period of time of entanglement revivals increases with the increase of $r$ and thus with the increase of the initial states purity.

\begin{figure}
\begin{center}
\includegraphics[width=4.25 cm, height=3.7 cm]{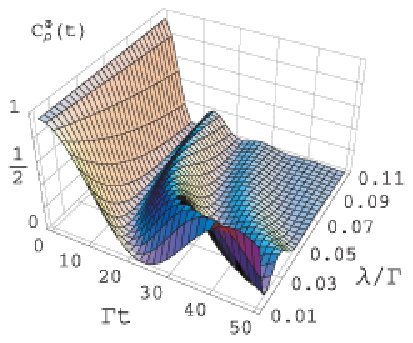}
\includegraphics[width=4.25 cm, height=3.7 cm]{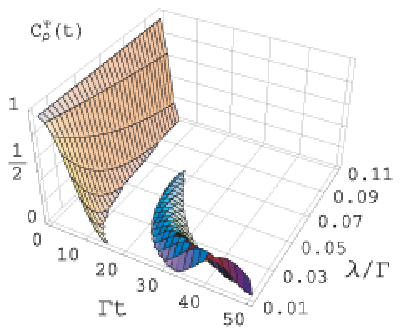}
\caption{\label{nc(T,t)}\footnotesize{(Non-Markovian case; $T=0$,
$r=1$, $\alpha^2=1/3$) Concurrence $C_\rho^{\Phi}(t)$ (left figure)
and $C_\rho^{\Psi}(t)$ (right figure) as a function of the
dimensionless quantities $\Gamma t$ and $\lambda/\Gamma$.}}
\end{center}
\end{figure}
Another aspect of interest is how the entanglement dynamics is influenced by the values of $\lambda/\Gamma$, which regulates the degree of non-Markovian characteristics of the system. This is an aspect that has never been examined previously. To this purpose we plot, in Fig.~\ref{nc(T,t)}, the concurrences $C_\rho^{\Phi}(t)$ and $C_\rho^{\Psi}(t)$ as functions of the dimensionless quantities $\Gamma t$ and $\lambda/\Gamma$, starting from pure Bell-like states, that is
for $r=1$ and taking $\alpha=1/\sqrt{3}$. This choice of the parameters
permits to evidence quite different time behaviors of the concurrences in terms of $\lambda/\Gamma$ for
the two initial Bell-like states $\hat{\rho}^{\Phi}(0)=\ket{\Phi}\bra{\Phi}$ and $\hat{\rho}^{\Psi}(0)=\ket{\Psi}\bra{\Psi}$ of Eq.~(\ref{istates}). The reason of the choice $\alpha=1/\sqrt{3}$ is that when $\alpha^2=1/2$, that is for initial maximally entangled pure states, the concurrences presents similar behaviors with the appearance of revivals of entanglement and dark periods but without the phenomenon of ESD. Instead, from Fig.~\ref{nc(T,t)} we see an evident difference in the entanglement dynamics of the two initial pure states. In particular, starting from the pure state $\hat{\rho}^{\Psi}(0)=\ket{\Psi}\bra{\Psi}$ with $\ket{\Phi}=(\ket{01}+\sqrt{2}\ket{10})/\sqrt{3}$, the left part of Fig.~\ref{nc(T,t)} shows neither ESD or dark periods of entanglement while the entity of the revival phenomenon increases by decreasing the coupling $\lambda/\Gamma$, that is by enforcing the non-Markovian effects. On the other hand, starting from the pure state $\ket{\Psi}=(\ket{00}+\sqrt{2}\ket{11})/\sqrt{3}$, the right part of Fig.~\ref{nc(T,t)} shows ESD occurrence for high values of $\lambda/\Gamma$, that is for small non-Markovian conditions, and revivals of entanglement after dark periods of time for sufficiently small values of $\lambda/\Gamma$, where the non-Markovian effects are stronger. The same analysis for mixed initial states ($r<1$) does not present a qualitatively different behavior of entanglement dynamics in terms of the amount of non-Markovianity regulated by the parameter $\lambda/\Gamma$.

These results summarize how the two-qubit entanglement dynamics in non-Markovian environment sensibly depends both on the initial conditions and the degree of non-Markovianity of the environments and show that different conditions may lead to very different time behaviors of the entanglement.

\section{\label{markoviandynamics}Markovian entanglement dynamics at $T\neq0$}
We now shall analyze in detail the two-qubit entanglement dynamics in Markovian environments at finite temperature, starting from the initial EWL states defined in Eq.(\ref{a}) and exploiting the well-known results obtained for the single-qubit reduced dynamics. In fact, on the basis of our procedure described in Sec.~\ref{procedure}, we shall obtain the reduced two-qubit dynamics by the knowledge of the reduced single qubit dynamics.

The single-qubit reduced dynamics associated to the Hamiltonian of Eq.~(\ref{Hamiltonian}) has been widely studied within Markov approximation for weak coupling. The single-qubit density matrix, $\hat{\rho}^S$, satisfies the
standard master equation in the Lindblad form \cite{barnbook}
\begin{eqnarray}
\frac{\textrm{d}\hat{\rho}^S}{\textrm{d}t}&=&\frac{\Gamma(\bar{n}+1)}{2}(2\sigma_-\hat{\rho}^S\sigma_+
-\sigma_+\sigma_-\hat{\rho}^S-\hat{\rho}^S\sigma_+\sigma_-)\nonumber\\
&+&\frac{\Gamma\bar{n}}{2}(2\sigma_+\hat{\rho}^S\sigma_--\sigma_-\sigma_+\hat{\rho}^S-\hat{\rho}^S\sigma_-\sigma_+),
\end{eqnarray}
where $\bar{n}=1/(e^{\frac{\hbar\omega_0}{k_BT}}-1)$ is the thermal
mean photon number of the resonant field mode at temperature $T$,
$\Gamma$ is the spontaneous emission rate and $\Gamma\bar{n}$
the stimulated absorption and emission rate. The solution of the
previous master equation can be found by solving the system of
differential equations associated to each matrix element of
$\hat{\rho}^S$ \cite{barnbook}. The reduced density matrix
elements can then be cast in the form of Eq.~(\ref{singlequbitevo}), with the
functions $u_t, v_t$ and $z_t$ given by
\begin{eqnarray}
u_t&=&\frac{e^{-\frac{\hbar\omega_0}{k_BT}}+e^{-\Gamma\coth(\frac{\hbar\omega_0}{2k_BT})t}(1-2e^{-\frac{\hbar\omega_0}{k_BT}})}{1-e^{-\frac{\hbar\omega_0}{k_BT}}},\nonumber\\
v_t&=&\frac{e^{-\frac{\hbar\omega_0}{k_BT}}}{1-e^{-\frac{\hbar\omega_0}{k_BT}}}(1-e^{-\Gamma\coth(\frac{\hbar\omega_0}{2k_BT})t}),\nonumber\\
z_t&=&e^{-(\Gamma
t/2)\coth(\frac{\hbar\omega_0}{2k_BT})-i\omega_0t}.
\end{eqnarray}
Substitution of these expressions in Eqs.~(\ref{rototdiag}) and
(\ref{rototnodiag}) allows to obtain the two-qubit density matrix at
time $t$ for both the initial states of Eq.~(\ref{a}). The concurrence is then obtained by using Eqs.~(\ref{concxstate})
and (\ref{concwstate}) as a function of time $t$ and the system parameters, that is purity $r$,
probability amplitude $\alpha = \sqrt{1- |\beta|^2}$ and temperature
$T$. We recall that, at finite temperature, X states have been shown in general to always
undergo sudden death of entanglement at finite time
\cite{qasimi}. In order to describe how the ESD time does depend on the physical
parameters of the system. In the following we shall plot the concurrence as a
function of $ \Gamma t$ and one of the dimensionless parameters
($r, \alpha^2, k_B T /\hbar \omega_0$), for assigned values of the remaining
ones.

\subsection{Low temperature}
\begin{figure}
\begin{center}
\includegraphics[width=4.25 cm, height=3.7 cm]{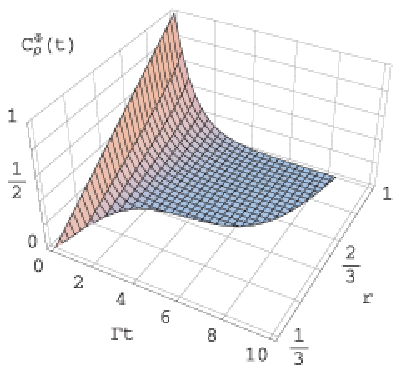}
\includegraphics[width=4.25 cm, height=3.7 cm]{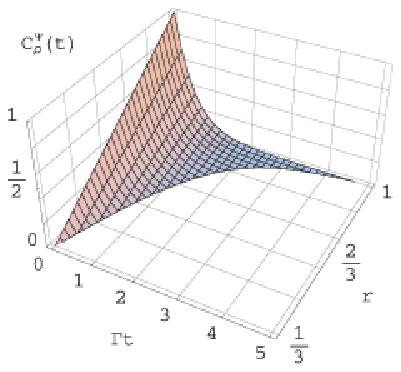}
\caption{\label{c(r,t)}\footnotesize{(Markovian case;
$\alpha^2=|\beta|^2=1/2$, $k_B T /\hbar \omega_0=0.1$) Concurrence
$C_\rho^{\Phi}(t)$ (left figure) and $C_\rho^{\Psi}(t)$ (right
figure) for Werner-like states as a function of the dimensionless quantities $\Gamma t$ and
the purity $r$.}}
\end{center}
\end{figure}
The concurrences $C_\rho^{\Phi}(t)$ and $C_\rho^{\Psi}(t)$ of
Eq.~(\ref{concwstate}), related respectively to the initial states
$\hat{\rho}^\Phi(0)$ and $\hat{\rho}^\Psi(0)$, are plotted in
Fig.~\ref{c(r,t)} as functions of the dimensionless time
$\Gamma t$ and purity $r$, for $\alpha^2=|\beta|^2=1/2$ and for a given low temperature such that $k_B T /\hbar\omega_0=0.1$. The choice of $\alpha$ corresponds, for the initial
states, to the Werner-like initial states, while the choice of low temperature
rests on the fact that in this limit the two-level approximation
for the atoms is more realistic. Fig.~\ref{c(r,t)} shows that the
ESD occurs for both states but at different times,
the differences depending on the values of purity $r$. In
particular, increasing the purity of the initial states, the ESD
time is retarded for both states. Moreover, the ESD time for the
initial state $\hat{\rho}^\Phi(0)$ is longer than the one for
$\hat{\rho}^\Psi(0)$. This may be linked to the fact that the product of populations of the totally excited state ($\rho_{11}(t)$) and ground state ($\rho{44}(t)$) of the two-qubit system can be shown to increase more slowly than the product of populations of the states where only one qubit is excited ($\rho_{22}(t)$, $\rho_{33}(t)$), as seen from Eq.~(\ref{concxstate}).

\begin{figure}
\begin{center}
\includegraphics[width=4.25 cm, height=3.7 cm]{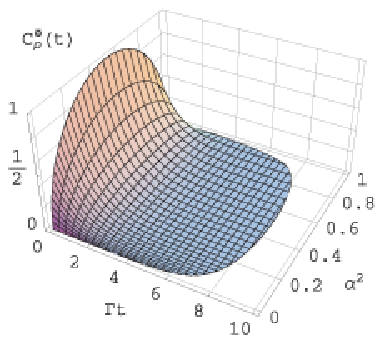}
\includegraphics[width=4.25 cm, height=3.7 cm]{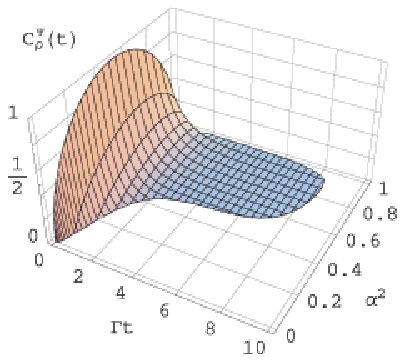}
\caption{\label{c(a,t)}\footnotesize{(Markovian case; $r=1$, $k_B T
/\hbar \omega_0=0.1$) Concurrence $C_\rho^{\Phi}(t)$ (left figure)
and $C_\rho^{\Psi}(t)$ (right figure) for Bell-like states as a function of the
dimensionless quantities $\Gamma t$ and $\alpha^2$.}}
\end{center}
\end{figure}
For $r=1$ the initial states reduce to Bell-like states, and we allow the degree of entanglement, represented by $\alpha^2$, to change again keeping $k_B T/\hbar \omega_0=0.1$. This differs from the $T\neq0$ cases studied previously, that were only restricted to initial Bell states \cite{carv1}. The concurrences $C_\rho^{\Phi}(t)$ and $C_\rho^{\Psi}(t)$ are then given in Fig.~\ref{c(a,t)}, which shows that ESD occurs for both initial pure states of Eq.~(\ref{istates}) independently on the initial degree of entanglement represented by $\alpha^2$. This is at variance with the zero temperature behavior where ESD occurs only for the initial pure state $\hat{\rho}^{\Psi}(0)=\ket{\Psi}\bra{\Psi}$ and when $\alpha^2<1/2$ \cite{santos}. The occurrence of ESD for both the initial pure states, when $T\neq0$ and independently from the degree of entanglement, is due to the population with time of the qubit excited states. However, in the $\hat{\rho}^{\Psi}(0)$ case it is the population of only one of the qubit, represented by the terms $\rho_{22}(t)$ and $\rho_{33}(t)$ in Eq.~(\ref{concxstate}), that makes after a certain time $K_2(t)\leq0$ in Eq.~(\ref{concwstate}). On the other hand, in the $\hat{\rho}^{\Phi}(0)$ case it is the excitation of both qubits, represented by the term $\rho_{11}(t)$ in Eq.~(\ref{concxstate}), that makes after a given time $K_1(t)\leq0$ in Eq.~(\ref{concwstate}). The right part of Fig.~\ref{c(a,t)} shows for $C_\rho^{\Psi}(t)$ an asymmetry in the times of ESD with respect to the maximally entangled case ($\alpha^2=1/2$), with shorter times for $\alpha^2<1/2$. This is again attributable to an initially higher contribution of the upper states for $\alpha^2<1/2$ for the state $\hat{\rho}^{\Psi}(0)$ of Eq.~(\ref{a}).

\begin{figure}
\begin{center}
\includegraphics[width=4.25 cm, height=3.7 cm]{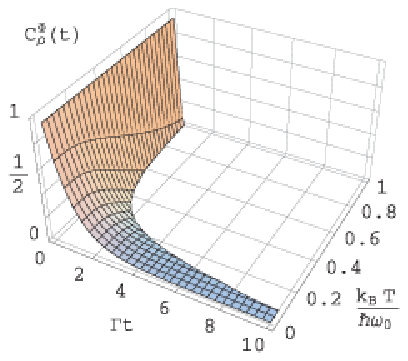}
\includegraphics[width=4.25 cm, height=3.7 cm]{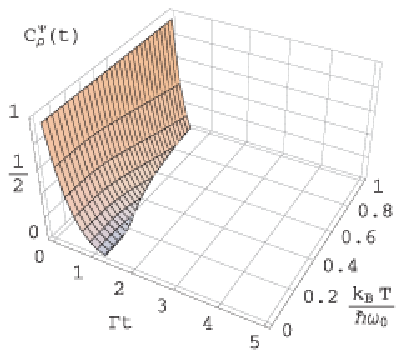}
\caption{\label{c(T,t)}\footnotesize{(Markovian case; $r=1$,
$\alpha^2=1-|\beta|^2=1/3$) Concurrence $C_\rho^{\Phi}(t)$ (left
figure) and $C_\rho^{\Psi}(t)$ (right figure) as a function of the
dimensionless quantities $\Gamma t$ and $k_B T /\hbar \omega_0$.}}
\end{center}
\end{figure}
In order to show how temperature influences the evolution of concurrence, in Fig.~\ref{c(T,t)} we plot the concurrences $C_\rho^{\Phi}(t)$ and $C_\rho^{\Psi}(t)$ as functions of the dimensionless quantities $\Gamma t$ and $k_B T /\hbar\omega_0$, fixing the purity ($r=1$) and the initial degree of entanglement ($\alpha^2 =1/3$). This choice of $\alpha$ allows to evidence a rather different behavior of concurrence for the two initial pure states $\hat{\rho}^{\Phi}(0)=\ket{\Phi}\bra{\Phi}$ and $\hat{\rho}^{\Psi}(0)=\ket{\Psi}\bra{\Psi}$ at small $T$. Moreover, for small values of $k_B T /\hbar\omega_0$, the ESD time is sensibly affected by slight temperature changes. Finally, for $T=0$ we also find that ESD occurs only for the pure entangled state $\hat{\rho}^{\Psi}(0)=\ket{\Psi}\bra{\Psi}$, as previously obtained \cite{santos}.

\subsection{High temperature}
The condition of low temperature $(k_B T/\hbar\omega_0\ll1)$ may result appropriate in some experimental contexts, as in the case of trapped atoms or ions \cite{roos-blattPRL2004,riebe-blattPRL2006}. However, there may be experimental situations where the low temperature condition cannot be applicable. This may be, for example, the case of two identical quantum dots each with energy difference between the first two levels given by $E_2-E_1\gg k_BT$. Under the action of an external magnetic field the ground state splits in two sublevels that may have an energy difference $\hbar\omega_0\ll k_BT$. In this case one may apply, in the study of entanglement between the sublevels of two qubits, the two-level approximation also in what for these sublevels is the high temperature regime.

In this high temperature limit $(k_B T/\hbar\omega_0\gg1)$, the concurrences of Eq.~(\ref{concwstate}) take a simple analytical form. Moreover, the concurrence $\widetilde{C}(t)$ takes the same form for both the two initial EWL states $\hat{\rho}^\Phi(0),\hat{\rho}^\Psi(0)$ of Eq.~(\ref{a}) given by $\widetilde{C}(t)=\textrm{max}\{0,\widetilde{K}(t)\}$ with
\begin{eqnarray}\label{conchigh}
\widetilde{K}(t)=2\left[\frac{r}{4}e^{-\frac{4\Gamma k_BT}{\hbar\omega_0}t}+r\alpha|\beta|e^{-\frac{2\Gamma
k_BT}{\hbar\omega_0}t}-\frac{1}{4}\right].
\end{eqnarray}
This expression is a generalization of the one obtained for pure maximally entangled states \cite{carv1} and takes into account mixedness and degree of entanglement of the initial states. It leads to a dimensionless ESD time $\Gamma\tilde{t}_\textrm{esd}$ whose expression in terms of temperature and initial degree of entanglement is
\begin{equation}\label{hightime}
\Gamma\tilde{t}_\textrm{esd}=-\frac{\hbar\omega_0}{2
k_BT}\ln\left(\sqrt{\frac{1+4r\alpha^2|\beta|^2}{r}}-2\alpha|\beta|\right).
\end{equation}
This expression shows that the dependence of $\tilde{t}_\textrm{esd}$ on the initial purity and degree of entanglement is weak. However, when $(r-r^\ast)/r^\ast\ll1$ $(r^\ast=\frac{1}{1+4\alpha|\beta|})$, that is for small initial entanglement, one gets
\begin{equation}\label{smallhightime}
\Gamma\tilde{t}_\textrm{esd}\approx\frac{\hbar\omega_0}{2 k_BT}[(1+4\alpha|\beta|)r-1]
\end{equation}
and the ESD time has a linear dependence on the purity $r$. When $r$ substantially differs from $r^\ast$, because of the weak dependence of $\tilde{t}_\textrm{esd}$ on $r$ and $\alpha$, we can limit to consider the case $\alpha=|\beta|=1/\sqrt{2}$ and $r=1$ when the dimensionless ESD time of Eq.~(\ref{hightime}) assumes the particularly simple form
\begin{equation}\label{ESDhightime}
\Gamma\tilde{t}_\textrm{esd}\approx0.88\frac{\hbar\omega_0}{2 k_BT} \qquad
(\alpha^2=|\beta|^2=1/2, r=1).
\end{equation}
In typical experimental conditions, quantum dots are subjected to an external magnetic field $B\sim1\ \textrm{--}\ 10\ \textrm{T}$ \cite{rinaldiPRL1996,kouwenhovenPRL2003}, inducing a sublevel splitting given by $\hbar\omega_0\approx\mu B$, where $\mu$ is the electron Bohr magneton, that corresponds to the dimensionless ESD time $\Gamma\tilde{t}_\textrm{esd}\sim(3\times10^{-1}\ \textrm{--}\ 3)/T$. For a temperature range $T\gg(0.3\ \textrm{--}\ 3) \textrm{K}$ we have $\tilde{t}_\textrm{esd}\ll1/\Gamma$. In this case, the limitation in the possible use of quantum gate operations with entangled quantum dots is determined by the ESD time instead that by the single quantum dot decoherence time. This lowers the number of operations $M$ \cite{barencoContPhys1996,divincenzoPRA1995} that the system can effectively perform.

\section{\label{concl}Conclusions}
In this paper we have presented a procedure (Sec.~\ref{procedure}) that allows to obtain the dynamics, represented by the time dependence of the total density matrix, of a system made by $N$ noninteracting qudits each under the influence of a local noise when the single qudit dynamics is known. This procedure is a generalization of the one developed previously for a system made by only two parts \cite{bellomo} and it is valid for any initial condition of the total system. Moreover, it results effective in simplifying the work typically required by the alternative approach where differential equations for the density matrix elements of the total system are solved \cite{qasimi,carv1}. It is also an useful tool in obtaining the corresponding dynamics for any part of the total system described by the reduced density matrix, as for example in obtaining the two-body dynamics from a three-body system. It remains of interest how to extend this procedure to the case when direct interaction between the systems is present \cite{tanas}.

Our procedure has been applied to a system of two identical independent qubits, each of them locally interacting with a bosonic reservoir. The general expression of the matrix elements of the reduced two-qubit system has been obtained in terms of three functions of time whose explicit form depends on the detailed properties of the reservoir but is independent on the initial state (Sec.~\ref{model}). These results have been used to study the entanglement dynamics of two identical qubits in a non-Markovian environment at zero temperature, taking as initial states the mixed EWL states defined in Eq.~(\ref{a}) (see Sec.~\ref{istatesandconcurrence}). This constitutes an extension of our previous results obtained for initial pure Bell-like states \cite{bellomo}. We have shown that the dynamics of entanglement, described by concurrence, presents revivals after periods of complete entanglement disappearance for some range of parameters, typically for high degree of entanglement of the initial states, as already found in the case of initial Bell-like states. Moreover, the phenomenon of ESD may occur depending on the values of parameters like purity or degree of entanglement of the initial state. Another parameter which is relevant in determining the entanglement dynamics is the degree of non-Markovianity in the system, that is represented by the ratio between environment effective spectral width and the spontaneous decay rate of the single qubit (Sec.~\ref{non-markoviandynamics}). The revivals of entanglement obtained for our two-qubit system differ from the ones previously obtained in the presence of interaction among qubits or because of their interaction with a common reservoir \cite{tanas,sun,hamdou}. The physical conditions examined here are similar to those typically considered in quantum computation and information, where qubits are independent and may interact with non-Markovian environments typical of solid state micro devices \cite{vega}. Our results show that also when initial states are not pure the occurrence of entanglement revivals can extend the time when entanglement can be usefully exploited.

We have also applied our procedure to the two-qubit system in a Markovian environment at finite temperature (Sec.~\ref{markoviandynamics}), again starting from extended Werner-like states, generalizing thus the results present in literature \cite{carv1}. We have shown that, in the low temperature case $(k_B T\ll\hbar\omega_0)$, ESD always occurs, at variance with the case of zero temperature where for some pure initial states it does not occur \cite{yu1,santos}. For qubits implemented by quantum dots the condition may occur that, instead, it is applicable the high temperature limit $(k_B T\gg\hbar\omega_0)$. Under this condition ESD always occurs and we have obtained a simple analytical expression of the ESD time in terms of the physical parameters such as purity and degree of entanglement of the initial state, that is a generalization of the results previously obtained for initial Bell states \cite{carv1}. In particular we have shown that, for some values of the temperature, the ESD time can be shorter than the single qubit decoherence time and this can give a limitation on the time of usefulness of entanglement in devices made by quantum dots.

\end{document}